\documentclass[twocolumn]{aastex631}

\usepackage{natbib}
\usepackage{hyperref}
\usepackage{rotating}
\usepackage{float}
\usepackage{graphicx}
\usepackage{longtable}

\shorttitle{Intracluster Light and Compact Sources in SMACS 0723}
\shortauthors{Martis et al.}

\newcommand\jwst{\textit{JWST}~}
\newcommand\hst{\textit{HST}~}

\begin{document}

\title{Modelling and Subtracting Diffuse Cluster Light in JWST Images: A Relation between the Spatial Distribution of Globular Clusters, Dwarf Galaxies, and Intracluster Light in the Lensing Cluster SMACS 0723}

\author[0000-0003-3243-9969]{Nicholas S. Martis}
\affiliation{NRC Herzberg, 5071 West Saanich Rd, Victoria, BC V9E 2E7, Canada}
\affiliation{Department of Astronomy \& Physics and Institute for Computational Astrophysics, Saint Mary's University, 923 Robie Street, Halifax, Nova Scotia, B3H 3C3, Canada}
\affiliation{Department of Mathematics and Physics, Jadranska ulica 19, SI-1000 Ljubljana, Slovenia}
\email{nicholas.martis@fmf.uni-lj.si}


\author[0000-0001-8830-2166]{Ghassan T. E. Sarrouh}
\affiliation{Department of Physics and Astronomy, York University, 4700 Keele St. Toronto, Ontario, M3J 1P3, Canada}

\author[0000-0002-4201-7367]{Chris J. Willott}
\affiliation{NRC Herzberg, 5071 West Saanich Rd, Victoria, BC V9E 2E7, Canada}

\author[0000-0002-4542-921X]{Roberto Abraham}
\affiliation{David A. Dunlap Department of Astronomy and Astrophysics, University of Toronto, 50 St. George Street, Toronto, Ontario, M5S 3H4, Canada}

\author[0000-0003-3983-5438]{Yoshihisa Asada}
\affiliation{Department of Astronomy and Physics and Institute for Computational Astrophysics, Saint Mary's University, 923 Robie Street, Halifax, Nova Scotia B3H 3C3, Canada}
\affiliation{Department of Astronomy, Kyoto University, Sakyo-ku, Kyoto 606-8502, Japan}

\author[0000-0001-5984-0395]{Maru\v{s}a Brada{\v c}}
\affiliation{University of Ljubljana, Department of Mathematics and Physics, Jadranska ulica 19, SI-1000 Ljubljana, Slovenia}
\affiliation{Department of Physics and Astronomy, University of California Davis, 1 Shields Avenue, Davis, CA 95616, USA}

\author[0000-0003-2680-005X]{Gabe Brammer}
\affiliation{Cosmic Dawn Center (DAWN), Denmark}
\affiliation{Niels Bohr Institute, University of Copenhagen, Jagtvej 128, DK-2200 Copenhagen N, Denmark}


\author[0000-0001-9414-6382]{Anishya Harshan}
\affiliation{University of Ljubljana, Department of Mathematics and Physics, Jadranska ulica 19, SI-1000 Ljubljana, Slovenia}




\author[0000-0002-9330-9108]{Adam Muzzin}
\affiliation{Department of Physics and Astronomy, York University, 4700 Keele St. Toronto, Ontario, M3J 1P3, Canada}

\author{Gaël Noirot}
\affiliation{Department of Astronomy and Physics and Institute for Computational Astrophysics, Saint Mary's University, 923 Robie Street, Halifax, Nova Scotia B3H 3C3, Canada}

\author[0000-0002-7712-7857]{Marcin Sawicki}
\affiliation{Department of Astronomy and Physics and Institute for Computational Astrophysics, Saint Mary's University, 923 Robie Street, Halifax, Nova Scotia B3H 3C3, Canada}

\affiliation{Niels Bohr Institute, University of Copenhagen, Jagtvej 128, DK-2200 Copenhagen N, Denmark}

\author[0009-0009-4388-898X]{Gregor Rihtar\v{s}i\v{c}}
\affiliation{University of Ljubljana, Department of Mathematics and Physics, Jadranska ulica 19, SI-1000 Ljubljana, Slovenia}


\begin{abstract}
We present a methodology for modeling and removing light from cluster galaxies and intracluster light (ICL) from \textit{James Webb Space Telescope} (\jwst) images of gravitational lensing clusters. We apply our method to Webb's First Deep Field  the SMACS 0723 Early Release Observations and use the ICL subtracted images to select a sample of globular clusters (GCs) and dwarf galaxies within the cluster. We compare the spatial distributions of these two samples with our models of the galaxy and ICL light, finding significant similarity. Specifically we find that GCs trace the diffuse ICL, while dwarf galaxies are centrally concentrated near the cluster center We quantify the relationship between the surface density of compact sources and total cluster light, demonstrating a significant, tight correlation. We repeat our methodology and compare distributions of GCs with dark matter surface density and find a comparable result. Our findings suggest a common origin for GCs and diffuse ICL, with stripping from massive galaxies as they merge with the cluster being a plausible scenario.

\end{abstract}



\section{Introduction} \label{sec:intro}
In the local universe, the study of the spatial distribution of globular clusters within galaxies \citep[e.g.,][]{Arakelyan_2018, Mackey_2019, Piatti_2019, de_Bortoli22} as well as in galaxy cluster environments \citep[e.g.,][]{durrell14, powalka18, madrid18} informs our knowledge of the formation history of the structures in which they are embedded. As some of the oldest objects in the universe \citep{Forbes_2018}, tracing the stellar properties of globular clusters can tell us how early star formation proceeded in these environments, whereas their locations and dynamics inform our knowledge of the dynamical history of their hosts. 

Detailed studies of the spatial distributions of globular clusters within a larger cluster environment have been carried out in the Virgo \citep{durrell14,powalka18} and Coma \citep{madrid18} clusters. The main findings of these studies have been that red (metal-rich) globular clusters are preferentially located around massive, early type galaxies, while blue (metal-poor) globular clusters have a more extended distribution in the intracluster regions. These intracluster globular clusters can form bridges between cluster galaxies, perhaps tracing interaction during infall of their host galaxies into the cluster.

The spatial resolution and sensitivity of the \hst combined with a significant investment of observatory time through the Hubble Frontier Fields program \citep{lotz17} enabled this study to be extended beyond the local universe for the first time by \citet{lee16} who analyzed the spatial distributions of globular clusters and dwarf galaxies in the Abell 2744 galaxy cluster. Thanks to the boost in sensitivity at infrared wavelengths offered by the \textit{James Webb Space Telescope} (\textit{JWST}), the detailed study of these objects beyond the confines of the local universe can now be achieved with modest observatory investment. Several authors have taken advantage of the early observations of the SMACS J0723.3–7327 (hereafter SMACS 0723) galaxy cluster at $z=0.39$ \citep{noirot23}, ``Webb's First Deep Field" to address this topic \citep{lee22,faisst22,diego23}. In this study, we extend this analysis by constructing a detailed model for cluster galaxy and intracluster light, finding a relation between the globular cluster number density and cluster surface brightness. Recently, \citet{harris23} have performed a similar analysis using \jwst observations of Abell 2744.

While the scientific focus of this paper is the study of the spatial distribution of compact sources, this paper also elaborates on the methods of image processing which enabled this analysis. This method has already been utilized by the CAnadian NIRISS Unbiased Cluster Survey (CANUCS) collaboration which has obtained data similar to the SMACS 0723 Early Release Observations. One of the primary goals of \jwst is to study the earliest galaxies to form in the history of the universe. Even with the unprecedented sensitivity of \textit{JWST}, observing such intrinsically faint targets will prove challenging. One strategy to overcome this challenge that has proved extremely fruitful with the \textit{Hubble Space Telescope} (HST) is to use the magnifying power of massive, lensing galaxy clusters, increasing both the sensitivity and spatial resolution of the telescope \citep{postman12,lotz17,coe19}. 
This boost in sensitivity offers a rare chance to study faint background sources in detail, but with the challenge of accounting for large, bright foreground galaxies ``in the way." In addition to the cluster galaxies themselves, one must contend with the intracluster light (ICL), composed of stars stripped from their host galaxies during interactions with the cluster environment \citep[see][for a review]{montes22review}. Previous studies have employed different strategies for modeling and removing this foreground light, including using parametric models from GALFIT \citep{peng10}, isophote modeling, and mode-filtering of the background \citep{merlin16, connor17, shipley18, lee22} and wavelet decomposition \citep{Livermore_2017}. We build on the strategy used in \citet{shipley18}, using isophote modeling combined with additional background subtraction to model cluster galaxies and the ICL. 

This paper is organized as follows. We introduce the data used in Section \ref{sec:data}, explain our cluster galaxy and background subtraction method in Section \ref{sec:methods}, describe our analysis of compact sources in Section \ref{sec:analysis}, present our findings in Section \ref{sec:results}, discuss these in Section \ref{sec:discussion}, and summarize in Section \ref{sec:summary}. Throughout this paper we assume a Flat $\Lambda$CDM Cosmology with $\Omega_\Lambda=0.7$ and $H_0=70$ km s$^{-1}$ Mpc$^{-1}$. 

\section{Data} \label{sec:data}
We use imaging data from the \jwst ERO program 2736 \citep{pontoppidan22}, which includes observations with {\em Near Infrared Camera} (NIRCam; \citealt{rieke23}) and {\em Near Infrared Imager and Slitless Spectrograph} (NIRISS; \citealt{doyon23}). NIRCam imaging was obtained in six broad-band filters: F090W, F150W, F200W, F277W, F356W and F444W. NIRISS imaging (collected to enable wide-field slitless spectroscopy) was obtained in the F115W and F200W filters. Estrada-Carpenter et al. (in prep.) demonstrates an extension of our cluster galaxy modeling strategy described below to slitless spectroscopy, but for the present work, we focus on the imaging data. These \jwst data are supplemented with HST/ACS imaging in F435W, F606W, and F814W from the RELICS program, drizzled to the same pixel grid \citep{coe19}.

We perform a custom reduction of the imaging data as described in \citet{noirot23}. This process utilizes a combination of a modified version of the \jwst pipeline and the \texttt{Grizli}\footnote{\href{https://github.com/gbrammer/grizli}{https://github.com/gbrammer/grizli}} (\citealp{brammer21}) grism redshift and line analysis software for space-based spectroscopy package. We obtain uncalibrated ramp exposures from the Mikulski Archive for Space Telescopes (MAST\footnote{\href{https://archive.stsci.edu/}{https://archive.stsci.edu/}}) and run a modified version of the \jwst pipeline stage Detector1, which makes detector-level corrections for ramp fitting, cosmic ray rejection, and dark current, and produces ``rate images." Our modified version of the pipeline also includes extra ``snowball" artifact flagging as well as a column-average correction for $1/f$-noise.  \texttt{Grizli} is used to align all exposures, subtract the sky background, and drizzle all images to a common pixel grid with scale 0\farcs04 per pixel. A secondary set of images with scale 0\farcs02 per pixel is generated for the NIRCam shortwave bands to be used only for source detection. The context for the \jwst Operational Pipeline (\texttt{CRDS\_CTX}) used for reducing the NIRISS (NIRCam) data was \texttt{JWST\_0932.pmap} (\texttt{JWST\_0916.pmap}). This is a pre-flight version of the NIRCam reference files, which we find does not accurately describe in-flight performance. We rederive photometric zero-points using \texttt{EAzY} \citep{brammer08} as described in \citet{mowla22}. The appropriateness of these corrections is verified by agreement of the photometric redshifts with spectroscopic redshifts from NIRISS slitless spectroscopy obtained by \citet{noirot23}.

\section{Cluster Galaxy and Background Subtraction} \label{sec:methods}

Before performing any modelling, we must first select galaxies to include in the cluster model. This was done by identifying bright cluster members by eye in color \hst images using the RELICS coverage of this field. Hereafter, we refer to the cluster galaxies included in our model as bright cluster galaxies or bCGs, as distinct from the central, brightest cluster galaxy denoted BCG as is common in the literature. In this nomenclature, the BCG is the brightest of our selected bCGs. Our selection is motivated by the desire to remove cluster light in order to study background galaxies rather than to create a full model of the cluster itself, so we focus on the largest and brightest galaxies rather than attempting to model all sources with redshifts consistent with the cluster. Our model includes 18 cluster galaxies whose properties are listed in Table \ref{table1}.

\begin{table*}
    \caption{Properties of bCGs included in the cluster model. Total fluxes for each model are given in $\mu$Jy. Note that galaxies 3 and 4 fall outside the NIRCam field of view, but are included for completeness since they are subtracted from the \hst imaging used to calculate photometric redshifts.} 
    \centering
    \begin{tabular}{cccccccccccccc}
    \hline
    \hline
ID & RA & DEC & F606W & F814W & F090W & F150W & F200W & F277W & F356W & F444W \\
0 & 110.82699500276941 & -73.45473834645486 & 43.4 & 189.7 & 205.2 & 598.5 & 812.7 & 1011.6 & 665.9 & 465.3 \\
1 & 110.79276728534073 & -73.447734701501 & 28.1 & 90.4 & 98.4 & 284.5 & 370.2 & 308.0 & 203.6 & 181.3 \\
2 & 110.80080009187155 & -73.44862372262301 & 19.5 & 69.3 & 91.2 & 257.9 & 348.6 & 317.3 & 206.4 & 180.4 \\
3 & 110.76997394775427 & -73.46911085243505 & 18.2 & 53.8 & --- & --- & --- & --- & --- & --- \\
4 & 110.80488047917918 & -73.48079823628191 & --- & 34.1 & --- & --- & --- & --- & --- & --- \\
5 & 110.81842414057762 & -73.45471268431392 & 12.4 & 44.6 & 69.9 & 222.1 & 279.4 & 322.2 & 192.3 & 115.4 \\
6 & 110.81869808054041 & -73.45532571056894 & 8.6 & 26.7 & 32.4 & 85.6 & 133.1 & 91.4 & 69.9 & 58.6 \\
7 & 110.85667189346512 & -73.44133548677209 & 9.5 & 29.1 & 37.7 & 105.5 & 136.1 & 120.1 & 92.6 & 73.9 \\
8 & 110.85399857467316 & -73.45016274374298 & 10.5 & 35.6 & 40.5 & 105.8 & 132.7 & 122.2 & 84.8 & 68.0 \\
9 & 110.87519968853212 & -73.45717367519062 & 10.7 & 28.7 & 38.4 & 126.2 & 159.8 & 146.2 & 96.1 & 79.0 \\
10 & 110.82588668112265 & -73.458790841949 & 8.0 & 25.8 & 33.2 & 87.1 & 112.5 & 89.6 & 67.3 & 55.2 \\
11 & 110.84581852555696 & -73.45144317672278 & 7.2 & 23.1 & 29.0 & 72.2 & 92.0 & 130.2 & 88.1 & 60.8 \\
12 & 110.82661540535587 & -73.45507671573814 & 1.7 & 4.3 & 8.5 & 49.8 & 10.5 & 17.9 & 2.5 & 8.7 \\
13 & 110.83778604894353 & -73.45628182975939 & 6.6 & 15.0 & 20.6 & 66.6 & 101.0 & 84.2 & 46.5 & 33.9 \\
14 & 110.855200969142 & -73.45029498431202 & 3.6 & 8.8 & 12.9 & 30.9 & 41.2 & 42.3 & 22.6 & 19.3 \\
15 & 110.83700826603199 & -73.45662063677356 & 3.1 & 14.3 & 15.6 & 35.1 & 46.4 & 41.6 & 28.0 & 24.8 \\
16 & 110.86979154739927 & -73.45014533161446 & 3.5 & 11.4 & 15.0 & 38.0 & 47.1 & 50.6 & 33.3 & 24.6 \\
17 & 110.84026397471291 & -73.45597299931083 & 8.4 & 26.1 & 29.4 & 78.2 & 97.1 & 98.5 & 61.7 & 51.1 \\
    \hline
    \end{tabular}
    \label{table1}
\end{table*}

Our strategy for generating bCG and ICL models is based on that described in \citet{shipley18} used to model and subtract bCGs from the \hst Frontier Fields. In brief, our method involves 5 steps after the initial selection of which galaxies to model: (1) initial background subtraction; (2) source segmentation; (3) mask refinement and initial isophote modeling; (4) model iteration; (5) final background subtraction. We describe these steps in detail below.

\subsection{Initial Background Subtraction}

As a preliminary step to the modeling process, the images are temporarily reprojected to a grid with scale 0\farcs08 per pixel. The lower resolution significantly quickens the modeling process and results in models of equivalent quality after reprojecting back to the original pixel scale as those produced obtained by performing the modeling at the original resolution. 

Before beginning to model individual galaxies, we use the \textsc{photutils} Background2D function to perform an initial subtraction of the ICL. We use a large box size of 20 pixels and filter size of 11 pixels at this stage to capture large-scale structure of the ICL as well as large defects remaining from the initial data reduction, such as imperfect background matching between regions covered by different NIRCam chips. This is important for the subsequent modeling because the isophote fitting algorithm attempts to remove all light from the region of the model, which can lead to substantial over-subtractions at large galactic radii. The initial background can also add light back in to account for obvious overdone background subtraction at the data reduction stage which resulted in regions of large negative flux. This is also important to correct because the isophote fitting algorithm will generate regions of negative intensity in order to bring the residual up to zero. 

\subsection{Segmentation}

We perform a preliminary source segmentation using \textsc{photutils} for two purposes. The first is to generate an initial mask for the isophote modeling. Since all sources other than the source to be modeled will be masked, we do not attempt to achieve optimal detection or deblending at this stage. The second purpose of the source detection is to obtain initial guesses for the morphological parameters of the isophote fits. The fitting algorithm is sensitive to the initial guesses, so these estimates must sometimes be manually adjusted, but we find them to be sufficient to obtain an initial model in the majority of cases. The size of each galaxy of interest measured at this stage is also used to determine the spatial extent of the model. We use the following parameters for this initial segmentation: source detection threshold of three times the initial background rms, minimum size of 20 pixels, convolution with a gaussian kernel of size 21 pixels, and a deblending contrast ratio of 0.02. 

\subsection{Initial Isophote Model and Mask Creation}

For the galaxy model creation we use the \textsc{photutils} elliptical isophote fitting tool. We begin by identifying the brightest cluster galaxy according to our source detection in the previous step and begin our modeling with this object. First, we create a cutout of the galaxy to be modeled and perform a fit using the segmentation map as a mask. Since the galaxies we model are quite extended in a crowded field, there are normally several sources within the spatial extent of the model. We perform another round of source detection on the residual image, which allows detection of smaller, fainter sources near the bCGs and better segmentation of other bright sources near the bCGs. This new segmentation image is then used as a mask to redo the isophote fitting for the current bCG. The model produced at this stage will be referred to as the ``iteration 0" model. This model is subtracted from the full image, removing its contamination from subsequent models before moving to the next brightest galaxy in our list. This process is repeated until we have an iteration 0 model for every bCG of interest. 

Due to the sensitivity of the fitting algorithm to initial guesses and variability of neighboring source density and brightness, we find it necessary to adjust many of the masks by hand. This task is mitigated by our decision to generate one mask for each instrument rather than for every filter. We find that when combined with sigma clipping of the input images, the masks generated this way are sufficient to obtain satisfactory models. The filters used to generate the masks are F814W for ACS, F200W for NIRCam, and F200W for NIRISS. 

\subsection{Model Iteration}

\begin{figure*}[ht!]
\includegraphics[width=\textwidth]{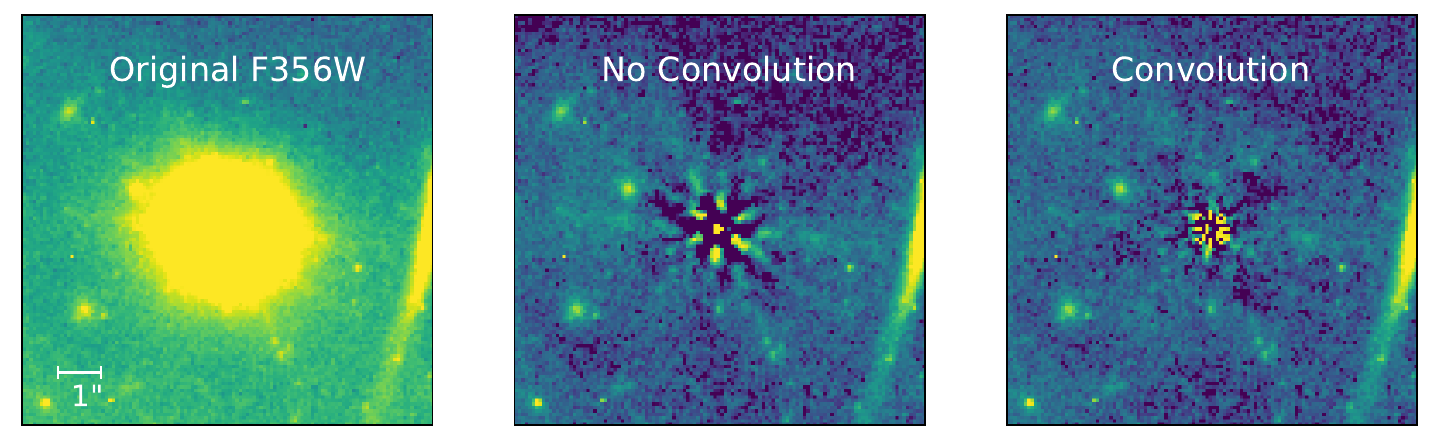}
\caption{Effect of including PSF convolution in the isophote modeling process. Left: original F356W image. Center: residual after subtracting a model without PSF convolution. Right: residual after subtracting a model with PSF convolution. This process significantly reduces the area contaminated by the residual, allowing reliable photometry of sources closer to the bCGs' centers.}
\label{fig:psf}
\end{figure*}

The model obtained at this stage is a good approximation of the cluster light contributed by large galaxies, but still not sufficient to produce subtracted images that can be used to detect small, faint sources in the background. First, any large residual may be misinterpreted as another source, particularly if the residuals are not symmetric around the bCG center. Second, we find that for bright sources, the isophote modeling leaves behind a positive residual showing the shape of the point spread function (PSF). These residuals contaminate a significant area (circular aperture of radius 2-3"), making photometry within this region impossible. Figure~\ref{fig:psf} shows an example of this PSF residual. This feature is not an effect of only \jwst imaging, but appears in deep \hst imaging as well, as may be observed in the bCG-subtracted images of the Hubble Frontier Fields in \citet{shipley18}. 

To improve the general quality of the models, we employ an iterative strategy. Again beginning with the brightest bCG, we re-insert the model back into the subtracted image. The galaxy is modeled again, but now with contaminating light from all other bCGs having been removed. This process produces a much better model in regions with significant overlap of bCG light. We continue through the list of bCGs until a new model has been produced for each, the sum of which becomes the new cluster model. We repeat this entire process ten times. The final bCG model is the median of these iterated models after removing the first iteration to allow a "burn-in." We use the median rather than the final model iteration because although the models generally improve with subsequent iteration, the fitting still occasionally fails. The median produces the smoothest residuals, which is our goal since we are interested primarily in background sources. 

This stage of the modeling also includes a prescription to account for the PSF residuals. After re-inserting the iteration 0 model into the image and producing a cutout that is used for the model, we deconvolve the cutout image using the \textsc{skimage} restoration package \citep{skimage}. We perform the isophote fitting on the deconvolved image, then convolve this model with the PSF. The resulting model includes the PSF feature that we observe in the iteration 0 residuals and effectively removes it. The result of this process is demonstrated in Figure~\ref{fig:psf}. As can be seen, this process enables reliable photometry much closer to the bCG center. 

\subsection{Final Background Subtraction}

\begin{figure*}[hb!]
\includegraphics[width=\textwidth]{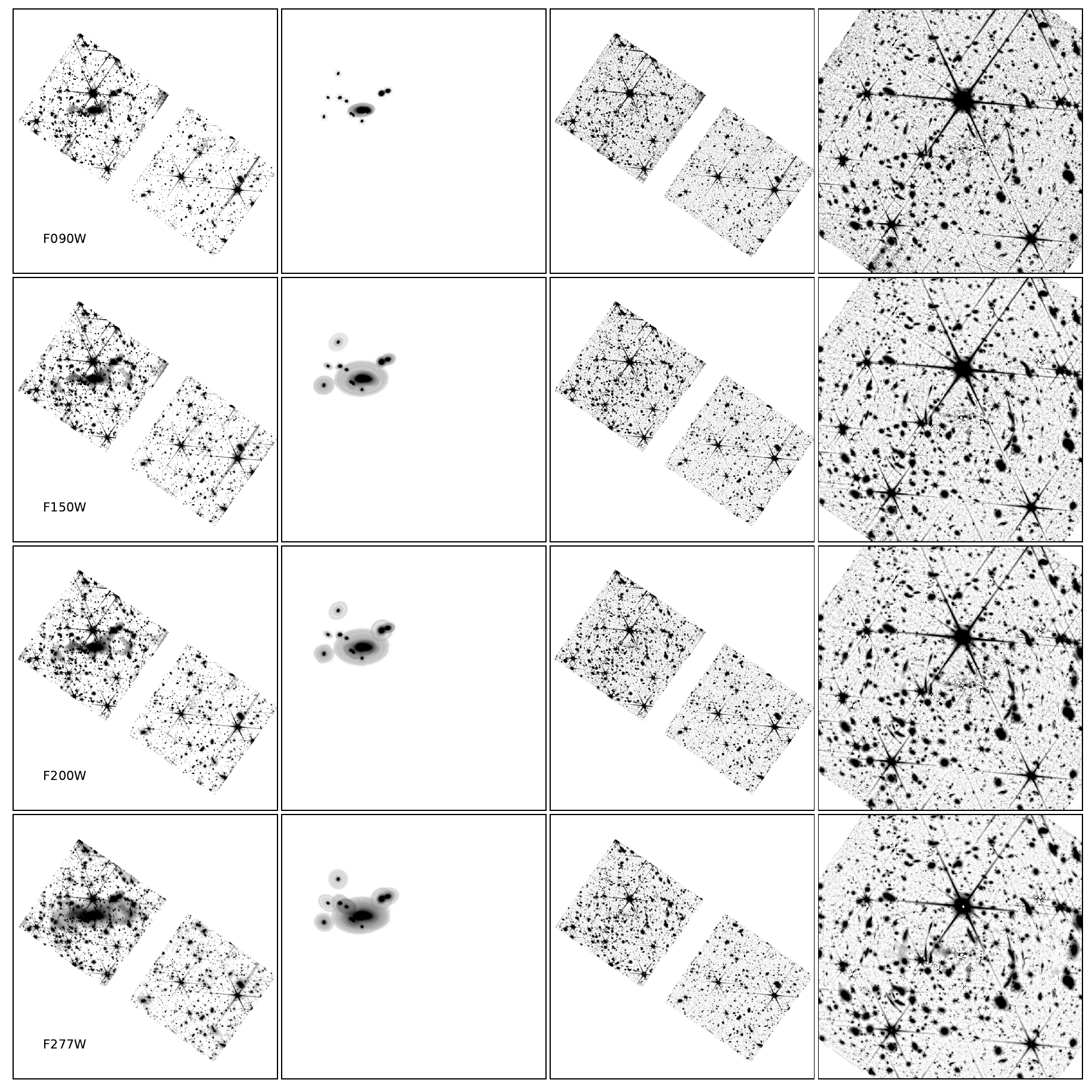}
\caption{Final results of our galaxy and background subtraction process. For each band we show the original image on the left, galaxy isophote model in the center left, cleaned image after subtracting the galaxy and background model on the center right, and a zoomed version of the cleaned image on the BCG region on the right. All images for a given filter are on the same brightness scale. NIRISS filters are indicated by the 'nis' suffix.}
\label{fig:main}
\end{figure*}

\begin{figure*}[ht!]
\includegraphics[width=\textwidth]{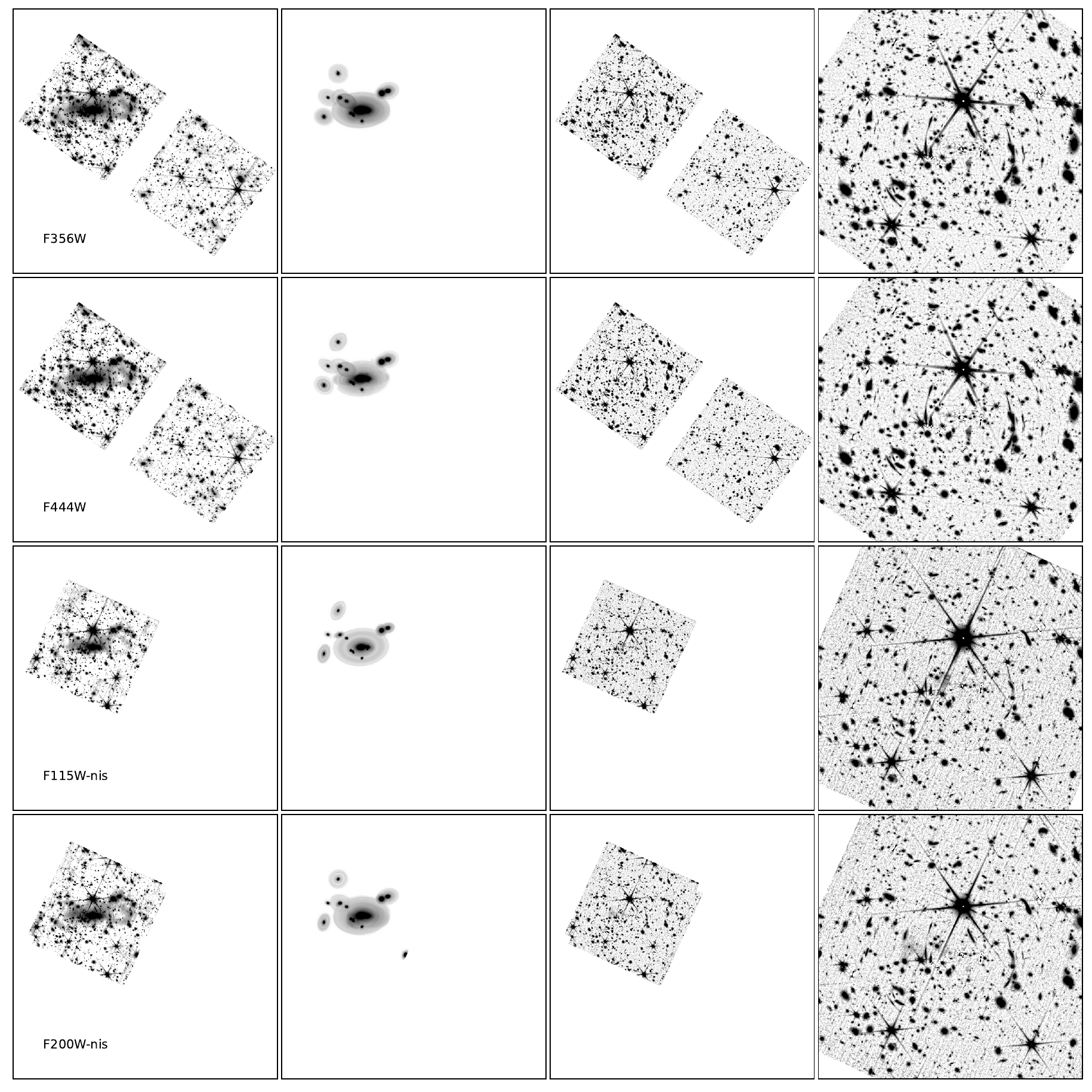}
\caption{Figure \ref{fig:main} continued.}
\label{fig:main-cont}
\end{figure*}

\begin{figure}[ht!]
\centering
\includegraphics[width=\columnwidth]{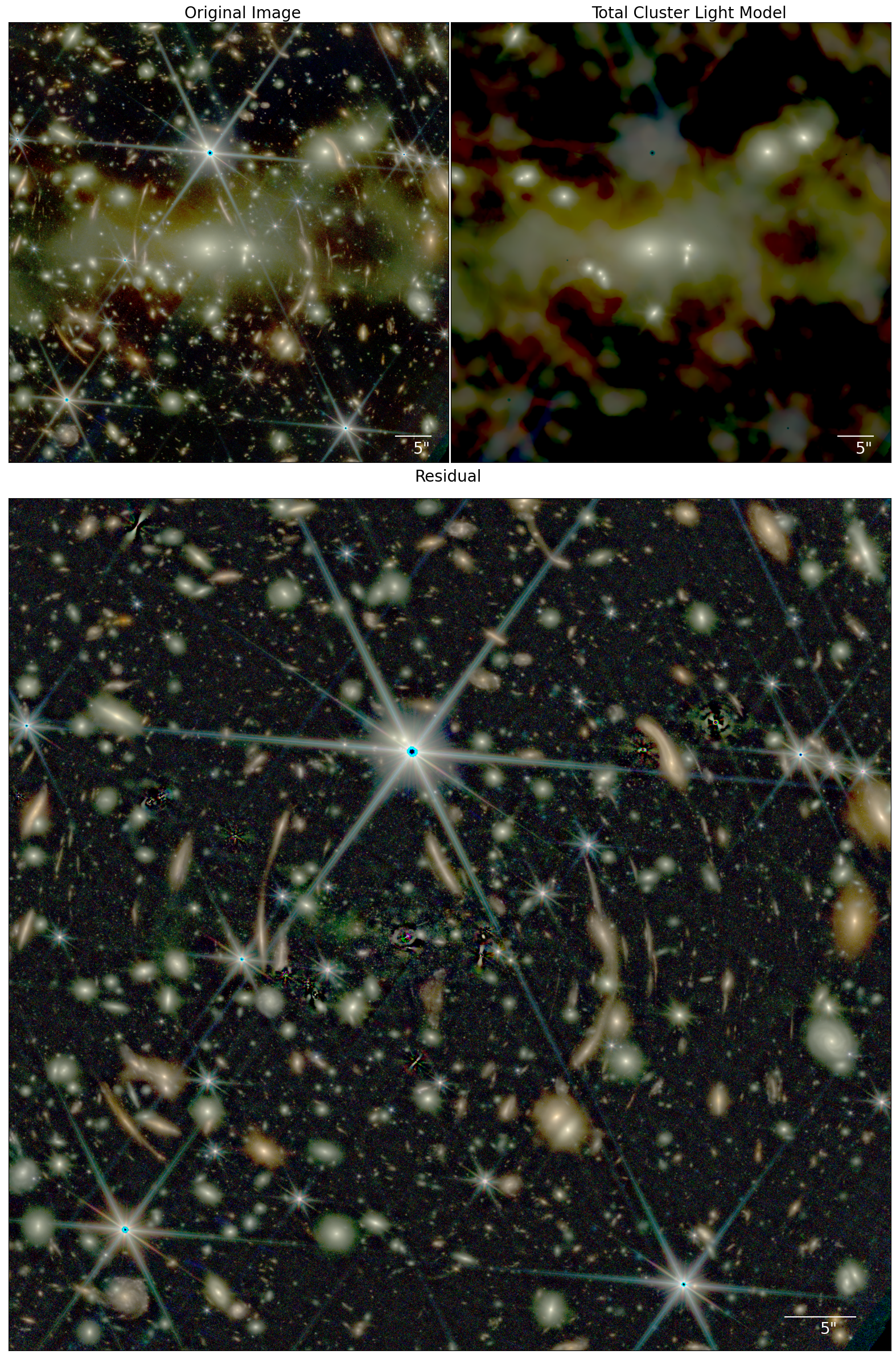}
\caption{Final results of our galaxy and background subtraction process shown as a composite image using all six NIRCam filters. The top left panel shows the original image, top right the total bCG+ICL model, and bottom the residual obtained by subtracting the model from the original image.}
\label{fig:color}
\end{figure}

After obtaining our final model of the bCGs themselves in the previous step, we interpolate the model back to the original pixel scale before subtracting it from the original image. We calculate a final background model with the goal of creating as smooth a background in the subtracted image as possible. At this stage we use much smaller box and filter sizes of 12 and 5 pixels respectively, thus allowing finer spatial variation of the background model. There exists an inherent trade-off in capturing as much of the ICL as possible and removing light from extended sources. Again, as we are primarily interested in faint sources we opt for a more aggressive subtraction. We also allow the background to take on negative values to account for any potential over-subtraction of the background at the data reduction stage. The output from this step is our final subtracted, science-ready image. The original images and final subtracted images for each of the NIRCam and NIRISS filters are shown in Figures~\ref{fig:main}, \ref{fig:main-cont}. The resulting color image with bCGs removed is shown in Figure~\ref{fig:color}.

\subsection{Caveats}
\begin{figure}[ht!]
\includegraphics[width=\columnwidth]{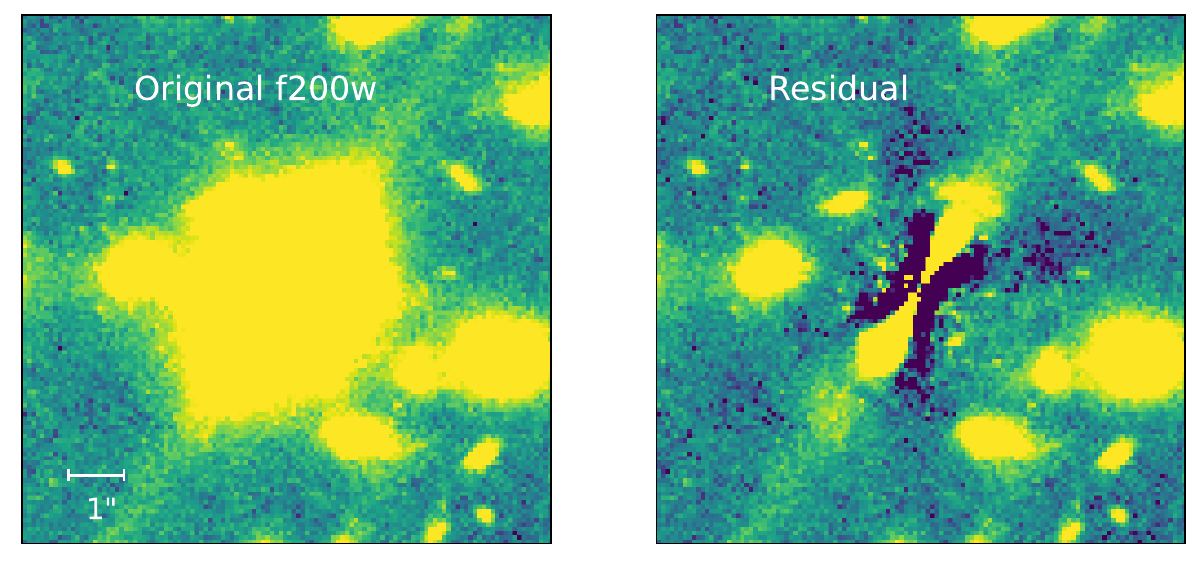}
\caption{Illustration of the limitations of elliptical isophote modeling on "disky" sources. The left panel shows the original NIRCam F200W image. The right panel shows the residual after subtracting the model, which exhibits the characteristic "peanut" shape. The residual image shows that while the model itself is unreliable near the bCG center, the subtraction still achieves the desired goal of decontaminating nearby sources.}
\label{fig:disky}
\end{figure}

The foreground light model presented in this paper enables study of high-z galaxies in the background with  a significant source of contamination removed. That being said, there are a few caveats arising from the chosen methodology that may be relevant to particular science cases. First, for those wishing to use the models themselves in order to study the cluster galaxies, it should be pointed out that high ellipticity sources are often not well-fitted by elliptical isophote models. This is a known issue and is related to the way the original \citet{jedrzejewski87} algorithm samples the isophotes (see \citealp{ciambur15} for a discussion). Figure \ref{fig:disky} shows the effect of this limitation in the present analysis. The resulting residual displays an under-subtraction along the major axis, and an over-subtraction along the minor axis. The large-scale diffuse light is still modeled well in this case, but we recommend using caution when performing photometry near the centers of the models. Because of this limitation, we have kept some high-ellipticity bCGs in the images because they could not be effectively modeled.

\begin{figure}[ht!]
\includegraphics[width=\columnwidth]{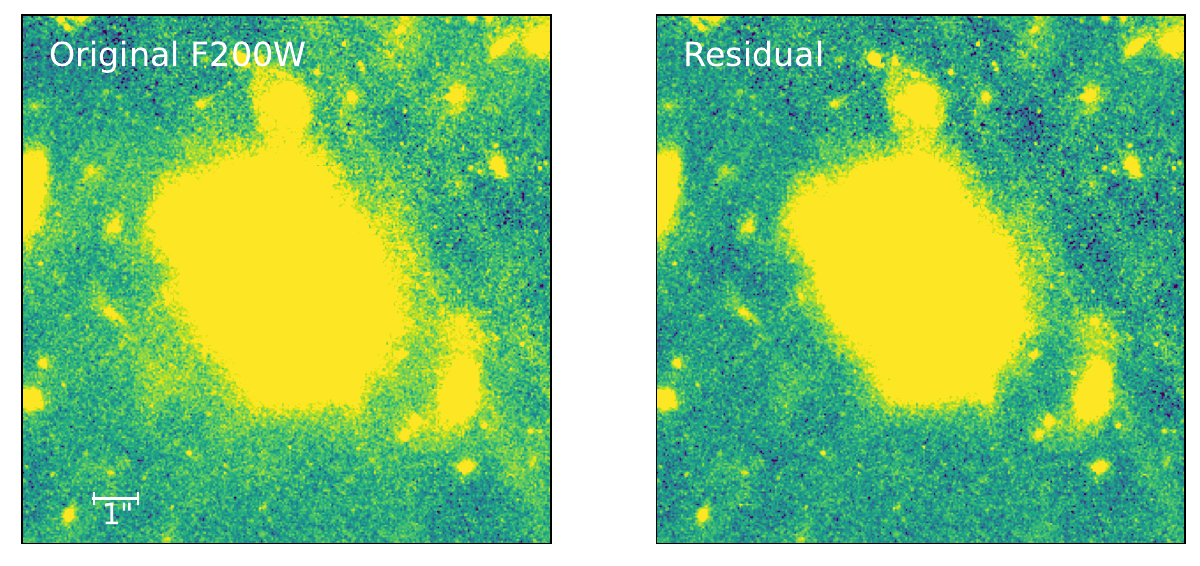}
\caption{Illustration of the effect of the aggressive background subtraction on large, extended sources. The left panel shows a cutout of the original image centered on an extended source, while the right shows the residual after the final background subtraction. Some light that is arguably part of the outer region of the source is subtracted.}
\label{fig:outskirts}
\end{figure}

Second, we note that some light may be removed from the outskirts of large, bright sources during our final background subtraction step. We have attempted to mitigate this issue by using a dilated source mask when calculating the final background, but there is no clear-cut boundary between the ICL and the extended light of large sources. A full measurement that aims to encapsulate all of either will encompass some contamination from the other. Figure \ref{fig:outskirts} illustrates this effect. We make our background available as a separate file for anyone wishing to use our cluster model with their own background treatment that may be more tailored to their particular science goal.

\section{Analysis} \label{sec:analysis}
We use the cluster model from the previous section to investigate the spatial distribution of compact sources in the SMACS 0723 cluster. We divide our analysis of compact sources into two classes, globular cluster (GC) candidates and cluster dwarf galaxies. Here we describe the photometric catalog construction and source selection. 

\subsection{Photometry}

To facilitate the detection of the faintest compact objects, we perform source detection at the finer 0.02" pixel scale to leverage the enhanced resolution of the NIRCam short wavelength channel. Source detection is performed on a co-added \texttt{CHI-MEAN} \citep{drlica-wagner18} image combining the F090W, F150W, and F200W filters. The resulting segmentation map is then regrided to the standard 0.04" pixel scale where photometry is performed across all 9 available \emph{JWST} and \emph{HST} filters in fixed circular apertures with diameters of 0.15", 0.3", 0.5", and 0.7". Total fluxes for all compact sources are measured in the 0.7" diameter aperture, and an aperture correction based on the F444W PSF is applied. We produce separate catalogs for each of the NIRCam modules, with the module centered on the cluster acting as our main sample, and the second module as a control/background sample. We verify the use of the second module as background below by looking at the distribution of compact sources across it. 

\subsection{Globular Cluster Selection}


\begin{figure}[ht!]
\includegraphics[width=\columnwidth]{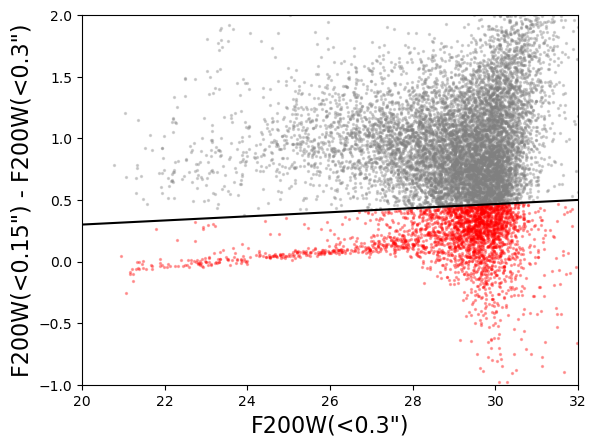}
\includegraphics[width=\columnwidth]{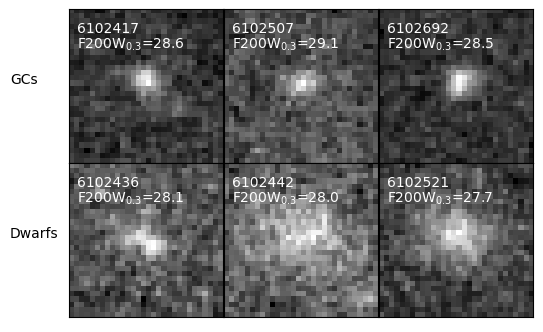}
\caption{Top: Difference in F200W magnitude in a 0.15" diameter aperture versus a 0.3" aperture as a function of F200W in the larger aperture for our full source catalog. Point sources form a sequence which was used to determine our selection for compact sources. Our point source selection is highlighted in red. Bottom: Examples of sources selected as compact, candidate GCs (top row) and extended, candidate cluster dwarf galaxies (bottom row). The ID number and 0.3" diameter aperture magnitude (uncorrected) for each source is provided. Image stamps are 0.6" on a side.}
\label{fig:point_source_selection}
\end{figure}

\begin{figure}[ht!]
\includegraphics[width=\columnwidth]{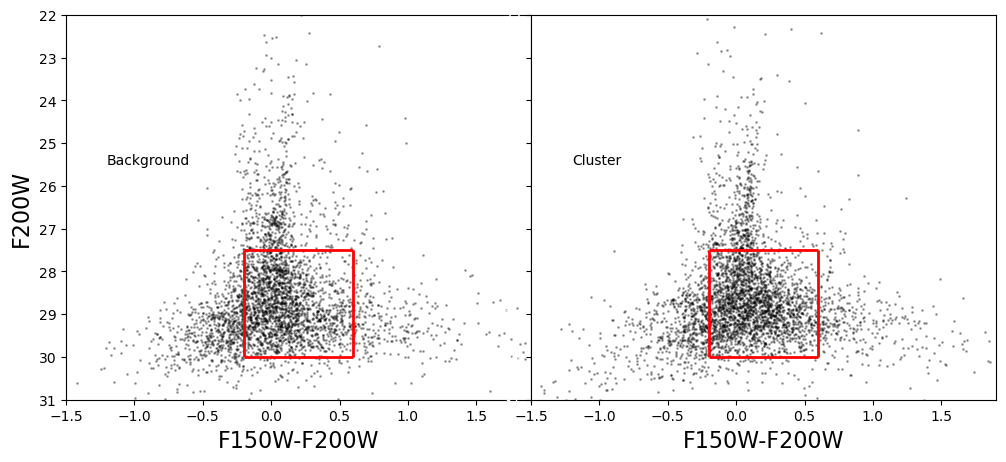}
\caption{F200W magnitude versus F150W-F200W color for our source catalogs of both NIRCam modules. The background module is shown on the left, and the cluster module on the right. Boxes indicate our color and magnitude selection for globular cluster candidates.}
\label{fig:color-mag}
\end{figure}

At the cluster redshift of $z=0.39$, globular clusters will be unresolved and detected as point sources given their small physical sizes of $\lesssim 10$ pc \citep{larsen01}. To identify point sources in our photometric catalog, we use a compactness threshold defined by the flux ratio in two different sized apertures, in which stars and other point sources form a clear linear sequence. We apply a selection of $F200W_{0.3"} > 0.983*F200W_{0.15"} + 0.033$ This selection is shown in Figure \ref{fig:point_source_selection}. Point sources clearly separate from extended sources up to $F200W \sim 28.5$ in this diagram, so we note that sources fainter than this limit are subject to uncertainty in the point source identification. Following \citet{lee22} we also apply cuts in F150W-F200W color and F200W magnitude corresponding to expected values for GCs at the observed redshift and in agreement with other studies of star clusters at in SMACS 0723 and Abell 2744 with \jwst. The criteria are $27.5 < F200W < 30$ and $-0.2 < F150W - F200W < 0.6$. This selection is shown for both catalogs in Figure \ref{fig:color-mag}. There is a visible overdensity of sources within the selection box in the cluster module compared to the background. This selection yields 601 and 311 objects in the cluster and background modules respectively. We note that the bright end of our selection may include ultra-compact dwarf galaxies in addition to GCs, as they are similarly unresolved \citep[see][]{janssens17}, so we refer to our sample as candidates throughout this work. A more detailed analysis of the stellar population properties may be able to disentangle these populations, but is beyond the scope of this work.

\subsection{Dwarf Galaxy Selection}
Using the photometric catalogs described above, we calculate photometric redshifts using the \textsc{EAZY} code. To select dwarf galaxies within the cluster, we choose objects with photometric redshift $0.1<z<0.6$ which meet the same color and magnitude cuts as the globular cluster selection, but not the point source selection. This results in 358 and 51 objects in the cluster and background modules respectively. Again, the significantly larger number of sources in the cluster module implies most of our selection is indeed associated with the cluster.

\subsection{ICL Model} \label{sec:icl}

\begin{figure*}[ht!]
\includegraphics[width=\textwidth]{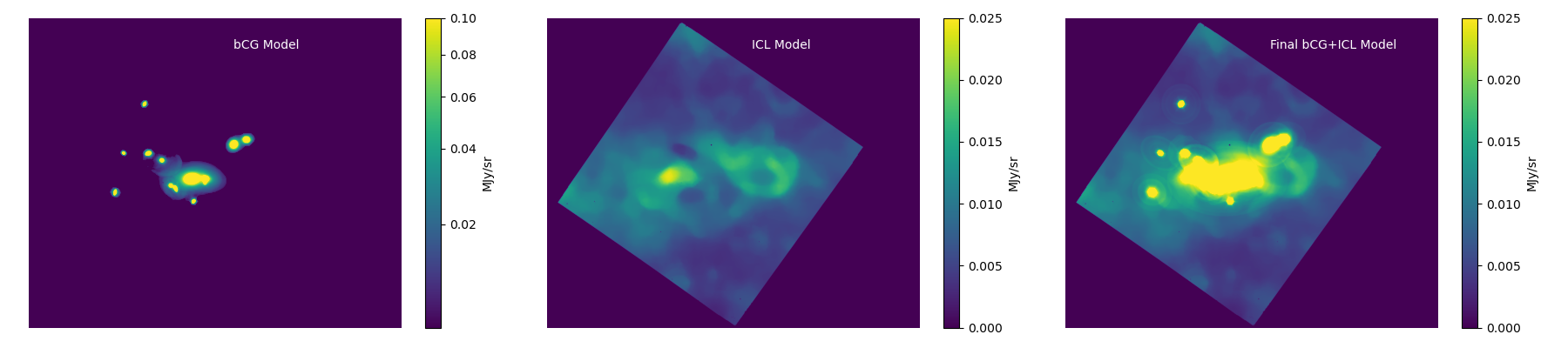}
\caption{Left: F277W bCG model. Center: ICL model generated using the process described in Section \ref{sec:icl}. Right: final bCG+ICL model.}
\label{fig:icl}
\end{figure*}

The primary goal of the cluster galaxy and ICL removal described in Section \ref{sec:methods} is to create clean images which can be used to obtain uncontaminated photometry. This process results in accurate models for the cluster galaxies, whereas the ``background" that is modeled out includes both the ICL and other sources of background remaining after earlier steps in the data reduction. In order to compare the spatial distribution of our compact sources with the ICL, we use the F277W NIRCam image to generate a model that better represents the ICL isolated from other background sources. We choose F277W for two reasons. One, the cluster galaxies and ICL have the highest signal to noise ratio in this filter, and thus stand above the other background components the most, allowing for a more reliable separation between them. At the redshift of the cluster, this corresponds to a rest-frame wavelength of 2$\micron$, which is close to the peak of light from an old stellar population. Two, the NIRCam longwave filter images cover the main body of the cluster with a single detector, mitigating potential zeropoint issues that may arise when calibrating the four detectors in the shortwave images. 

Rather than using the fully reduced images for the ICL model, we generate a drizzled image with all background subtraction turned off. The first step is to remove the zodiacal background, which is done by subtracting $95\%$ of the median pixel value with sources masked. To generate the ICL model, we iteratively apply the \textsc{ASTROPY} Background2D function on smaller spatial scales using the median background estimator. Specifically, we apply two passes, the first with a box size of 80 pix and filter size of 29 pix, the second with a box size of 25 pix and a filter size of 9 pix. The initial background subtraction accounts for the large-scale ICL structure whereas the second captures finer features. The final model is restricted to have flux values greater than zero to remove the artifact of slight over-subtraction in some regions of the bCG model. We note the final ICL model contains nonphysical background artifacts, most noticeable in the northern and eastern corners of the image, but at a surface brightness level significantly lower than the main ICL features (not more than $30\%$). We note that we exclude from our source detection and subsequent analysis the image edges due to the higher background noise, so this limitation does not affect our results.  The final ICL model is combined with the galaxy model and shown in Figure \ref{fig:icl} to obtain a total cluster light model.

\section{Results}\label{sec:results}

\subsection{ICL Structure}

Figure \ref{fig:icl} shows the F277W bCG model, ICL model, and final bCG+ICL model. We recover several features described in previous work including a stream on the eastern side and large loop to the west of the bCG \citep{montes22}. Removing the bCGs also makes apparent faint shell or stream structures ringing the center of the cluster. The combined bCG+ICL model shows that the total cluster light is aligned with the axis of the bCG in the East-West direction. \citet{montes22} infer the central region to be formed by a major merger, whereas the outer region is composed of tidally stripped remnants of satellites.

\subsection{Spatial Distribution of Compact Sources}

\begin{figure*}[ht!]
\includegraphics[width=\textwidth]{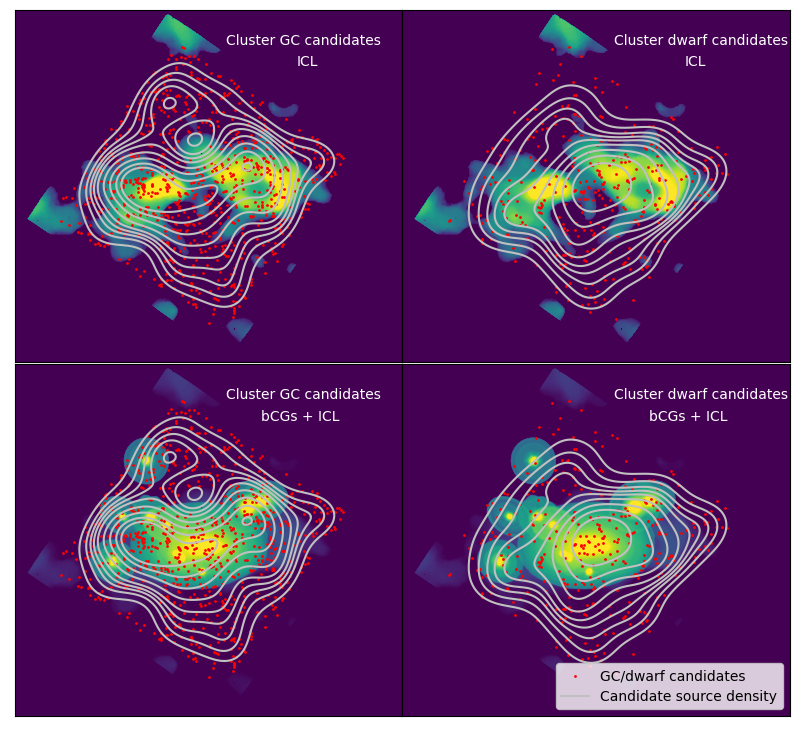}
\caption{Spatial distribution of our globular cluster candidates and cluster dwarf galaxies overlaid on our  intracluster light model (top two panels) and total cluster light model (bottom two panels) (see Section \ref{sec:results}). Density contours for globular cluster and dwarf galaxy candidates are shown at ten equally spaced levels. }
\label{fig:points_over_models}
\end{figure*}

Here we investigate the spatial distributions of our two samples with respect to both our ICL and bCG+ICL models. Figure \ref{fig:points_over_models} shows the spatial distributions of our GC and dwarf cluster galaxy candidate selections overlaid on the intracluster (top two panels) and total cluster light (bottom two panels) models. Contours designate ten evenly spaced density levels. In all cases, the distributions appear to correspond to the shape of the cluster light. By eye, the GC candidates appear to more closely follow the ICL, with the highest density appearing over the eastern lobe and a significant overdensity around the western loop. In contrast, the dwarf galaxy candidates appear to more closely follow the total cluster light model, with the highest density occurring at the position of the bCG and roughly following its shape. 

\begin{figure*}[ht!]
\includegraphics[width=\textwidth]{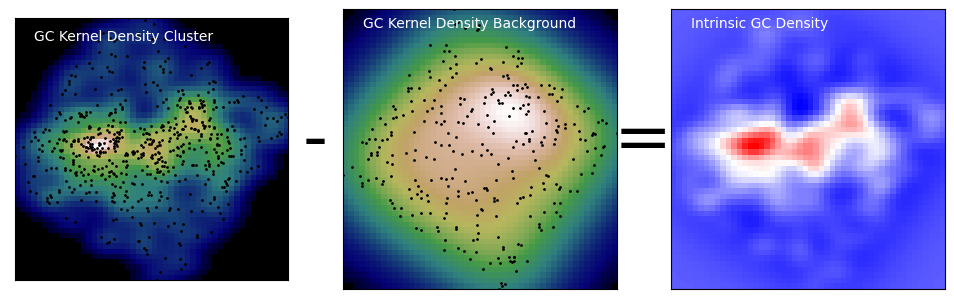}

\includegraphics[width=\textwidth]{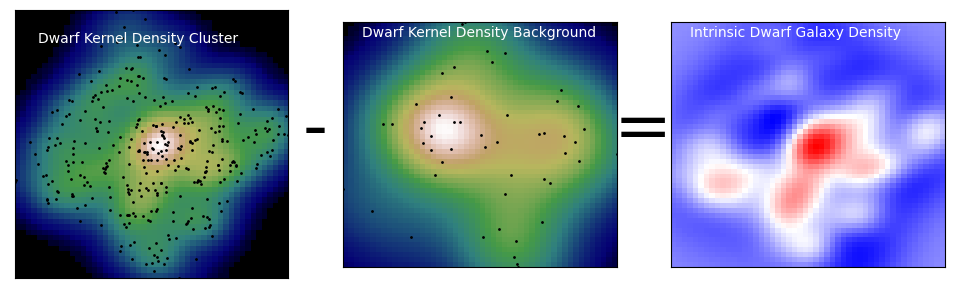}
\caption{Top Left: Kernel density for spatial distribution of GC candidates in the cluster module. Positions of individual sources are marked as black points. Top Center: Same for background module. Top Right: Excess kernel density of GC candidates in the cluster module, computed by subtracting the median density of the background module. This gives the intrinsic distribution of GC candidates without contamination from background point sources which meet the source selection such as distant galaxies. Bottom Row: as above for dwarf galaxy candidates.}
\label{fig:kernel_density}
\end{figure*}

To test this hypothesis, we use a gaussian kernel density estimator to calculate the intrinsic 2D surface density of the spatial distribution of GC candidates. Distant galaxies or foreground stars may also appear as point-like sources that mimic GCs in the imaging, leading to an overestimation of their surface density, so this process removes that effect. The top row of Figure \ref{fig:kernel_density} shows our results, with the cluster module in the left panel, and the background module in the center panel. The left panel contains the information shown by the contours in Figure \ref{fig:points_over_models}. We can see that the density kernel of the background field (middle panel) is comparatively uniform within the region covered by the NIRCam field of view. This is what we expect for a random selection of objects not associated with the cluster. In contrast, the  right panel shows the excess kernel density of GC candidates in the cluster module, computed by subtracting the median density of the background module. This removes the effect of objects which we do not expect to be associated with the cluster from our density estimation. The resulting map shows significant structure that closely resembles the ICL model shown in the top panels of Figure \ref{fig:points_over_models}. Specifically, we see over-densities of GCs in the region of the eastern lobe and western loop of the ICL, particularly its northern half.

Similarly, the bottom row of Figure \ref{fig:kernel_density} shows the same process for our dwarf galaxy candidate selection. As before, the background module in the center panel shows a uniform distribution, as expected for a random selection of sources. As suggested in Figure \ref{fig:points_over_models}, after subtracting the background level given by the background module, we observe a strong over-density near the position of the BCG. We also observe a weaker over-density extending to the collection of cluster galaxies to the southeast. 


\subsection{Correlation Between ICL Brightness and Compact Source Density}

\begin{figure}[h]
\includegraphics[width=\columnwidth]{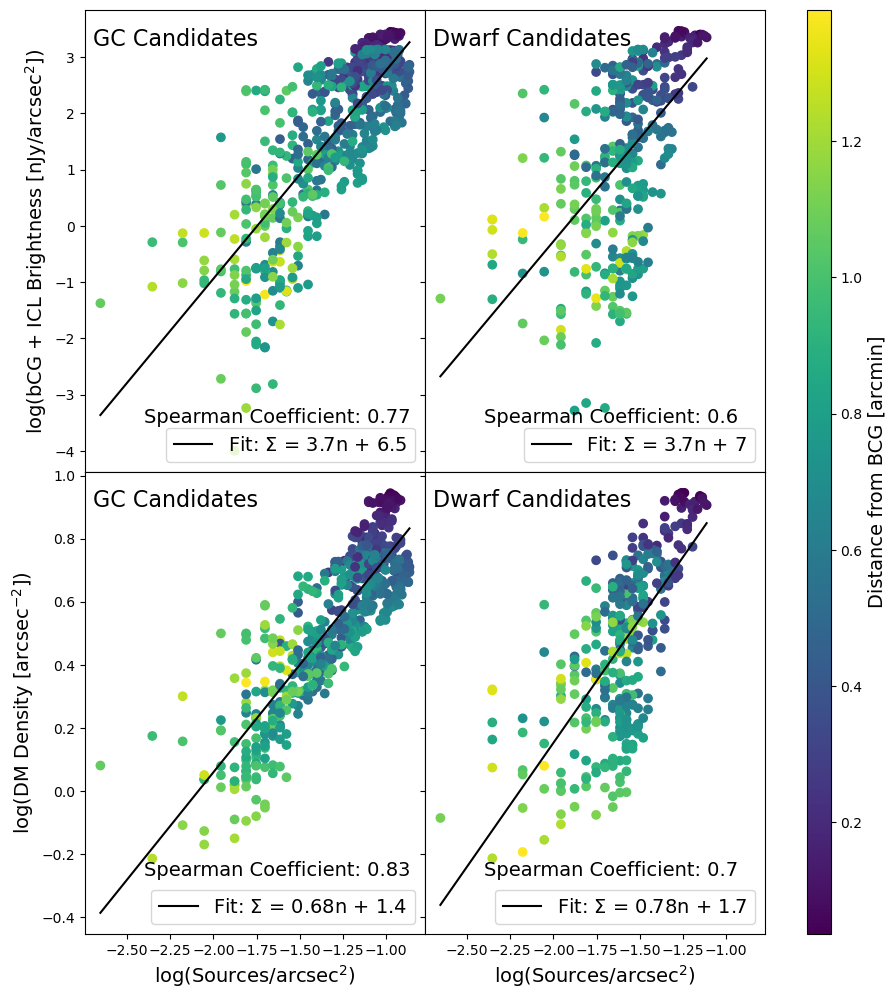}
\caption{Top: Surface brightness of the total cluster light model versus on-sky density of GC candidates (left) and cluster dwarf candidates (right). Bottom: Scaled surface mass density of dark matter \citep[derived from the convergence map from][]{mahler23} versus sky density for our two samples as above. Points are colored by distance from the center of the BCG. Each panel shows the  result of a power law fit to the relation as well as the Spearman correlation coefficient between the two quantities. }
\label{fig:sb_vs_n}
\end{figure}

The top two panels of Figure \ref{fig:sb_vs_n} show the surface brightness of our total cluster light model versus the number density of compact sources for our GC (left) and cluster dwarf (right) samples measured in a 12" aperture. Points are colored by their distance from the center of the BCG. There is a clear correlation between both the GC and dwarf candidate source density and surface brightness of cluster light. The coloring indicates that the regions of highest density correspond to the center of the cluster defined as the center of the BCG. We fit a linear relation between these quantities in log-log space, such that $\Sigma = \alpha n + \beta$ where $\Sigma$ represents the logarithm of surface brightness or surface density, and n is the logarithm of the 2D number density of sources. We display the results of the fits in each panel. The slope of the relation between source density and cluster light surface brightness is $\sim 3.7$ for both GC and cluster dwarf candidates. Each panel also shows the Spearman correlation coefficient between the two quantities. We find a stronger correlation between the GC candidate sources and the cluster light than for dwarf galaxies. We point out that the correlation spans over three orders of magnitude in surface brightness of cluster light, from faint ICL features (in the bottom left of the plot), to the center of the BCG (dark points at the top right of the plot).

It is also reasonable to hypothesize that the spatial distribution of compact objects in the cluster will be associated with the distribution of mass, which is imperfectly traced by the cluster light. To test this possibility, we also compare the surface density of our two samples to the dark matter surface density obtained from the convergence map presented by \citet{mahler23}. The results are shown in the bottom two panels of Figure \ref{fig:sb_vs_n}. We find strong correlations between the dark matter and source density of both our samples, with the correlation for GC candidates again being stronger. In this case we find slopes of 0.7 for the GC candidates, and 0.8 for the dwarf galaxies. Thus the slope of the relation of source density with cluster surface brightness is higher than that with dark matter, covering a wider dynamic range.

\begin{figure}[h]
\includegraphics[width=\columnwidth]{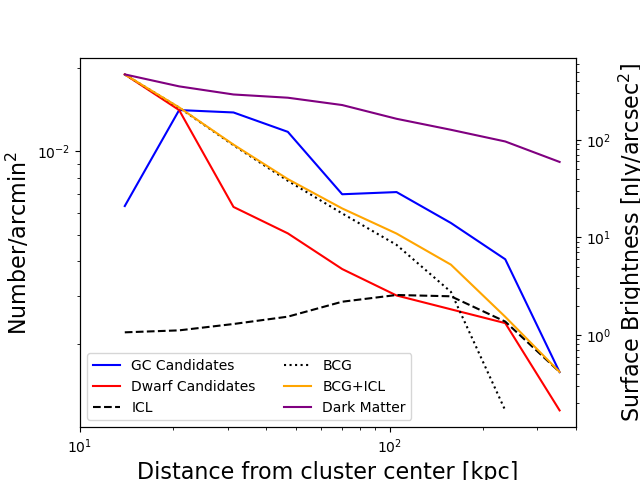}
\caption{Density profiles of the GC (blue) and dwarf galaxy (red) candidate selections. Also shown are the surface brightness profile of the central BCG (dotted curve) and ICL (dashed curve) as well as their combination in the total cluster light model (yellow). The dark matter density profile (purple) is scaled arbitrarily to match the range of surface brightness of the cluster light model.}
\label{fig:profiles}
\end{figure}

Finally, in order to facilitate comparison with previous work, we compare the density profiles of our two samples with the ICL and BCG surface brightness, and dark matter density profiles. These are shown in Figure \ref{fig:profiles}.

\section{Discussion}\label{sec:discussion}

\subsection{Correlation of Compact Sources with Cluster Structure}
Observations of the Virgo and Coma clusters have clearly demonstrated the GC populations in galaxy clusters extend well beyond the vicinity of cluster member galaxies into the intracluster medium and trace ICL \citep{durrell14,madrid18}. These findings were extended to higher redshift and cluster mass by \citet{lee16}, who examined the spatial distribution of GCs and compact dwarfs in the Hubble Frontier Fields cluster Abell 2744 at $z \sim 0.3$. The correlation between GC and ICL surface density suggests both are subject to the same dynamical forces within the cluster and possibly point to a common origin. In SMACS 0723, the fact that they are currently closely physically associated and have not dynamically diverged may suggest they have been stripped from their host galaxies only recently. The velocity offset of the BCG with respect to the systematic velocity of the cluster \citep{mahler23} and the highly asymmetric structure of the ICL are both indicative of ongoing merging processes. Together, these observational signatures may suggest that the ICL originates preferentially from stripping of the halos of massive galaxies along with their GCs, rather than from the tidal disruption of dwarf galaxies. 


\citet{lee22} have performed their own analysis of the spatial distribution of GCs in SMACS 0723. They find the GC spatial distribution to correlate with both the BCG and diffuse light in the cluster. The overall distribution is peaked near the center of the BCG and elongated along its major axis. Overdensities of globular clusters are also co-spatial with other bright cluster galaxies. A concentration of intracluster globular clusters coincides with the western loop of the ICL. From a purely visual inspection, they find the distribution closely follows the structure of the dark matter mass map from \citet{mahler23}. Figures \ref{fig:points_over_models} and \ref{fig:kernel_density} show that we recover the same qualitative behavior. We have extended this analysis by calculating the quantitative relation between these quantities. 

Recently, \citet{harris23} have performed a similar analysis using publicly available NIRCam mosaics of Abell 2744 at z = 0.308. They find about 10,000 point-like objects likely to be globular clusters, preferentially distributed near the five largest cluster galaxies. The ICL of this cluster is concentrated around these five galaxies, so we obtain a similar result: GCs trace cluster light rather than the center of the cluster mass profile. \citet{liu20} studied the properties and spatial distribution of ultra-compact dwarfs in the Virgo cluster and found that they are predominantly located near the brightest galaxies. \citet{harris23} find a similar result in Abell 2744, such that the brightest objects in their sample, which are likely to be ultra-compact dwarfs, are almost universally located near bright galaxies, while fainter sources can be found in the intergalactic regions. A detailed study of the stellar masses and metallicities of these objects, similar to that presented in \citet{faisst22} may help to address this hypothesis.  

Since we have light profiles for each of our modeled galaxies, it is trivial to determine the fraction of sources that fall within the extent of our models and are thus likely associated with bCGs. We find 49\% of GC candidates and 41\% of dwarf candidates lie within twice the half-light radius of a bCG. This is lower than is typically found in the local universe with the Virgo cluster having 82\% \citet{durrell14}, and  Coma having 55-70\% \citet{peng11}. Using JWST data, \citet{harris23} find 88\% of GCs lie close to bright galaxies in Abell 2744. Interestingly, the two clusters with the most GCs in the intergalactic medium, SMACS 0723 and Coma, both have significant ICL features not associated with individual galaxies. This again points to the diffuse light and GCs issuing from a common origin, with the ICL and GCs likely being stripped from the outskirts of massive galaxies due to ongoing merger activity in the recent past, and observation occurring before the diffuse light and GCs can settle onto the BCG.

In contrast, our dwarf galaxy sample is centrally concentrated on the BCG.  Extended objects with small masses are unlikely to survive the gravitational potential of the cluster center for long, meaning that perhaps tidal interactions will remove the envelopes of some of these sources leaving compact remnants behind in orbits around the BCG and southeastern galaxy group. Thus we may be observing the precursors of ultra-compact dwarfs observed in the close vicinity of bCGs noted by the above studies.


We find that the GC spatial distribution correlates more strongly with both the cluster light and dark matter distributions than the cluster dwarf population. From Figure \ref{fig:sb_vs_n}, the respective GC and dwarf relations appear to exhibit comparable scatter, but our sample of GCs is about twice as large, meaning the stronger correlation may be observationally driven. With our current methodology, our source detection remains robust beyond the cluster-centric distance which we are able to reliably measure the ICL as its brightness approaches the background level of the images, introducing noise into the measurement. In the lower density regime, the correlations with dark matter appear to have less scatter. However, it is important to note that the dark matter model extends beyond our cluster light model, leading to a more reliable measurement in this regime. This is partially driven by the fact that the dark matter density profile is shallower than the cluster light profile, but the dark matter model also includes a number of halos at greater cluster-centric radii than the galaxies which we model. The present work includes 18 cluster galaxies, but our models for the five CANUCS clusters contain of order 40-70 galaxies. Future work will demonstrate whether the increased scatter in these relations at the low density end is driven by the low number of cluster galaxies modeled here.  

\subsection{Density Profiles}
\citet{lee16,lee22,diego23} find in observations of Abell 2744 and SMACS 0723 that intracluster GCs have a steeper number density profile than the dark matter mass profile. The same has been replicated in simulations. In nine simulated galaxy clusters from Illustris \citep{ramos-almendares20}, ICL and intracluster globular clusters have a density profile scaling of $\sim r^{-2.5}$, which is steeper than the dark matter profile in agreement with previous observations of both local clusters and SMACS 0723 as well as our own. \citet{alonso_asensio20} analyze the EAGLE simulations and find similar results. Physically, this points to a scenario in which globular clusters undergo more significant stripping from their hosts as the gravitational potential becomes stronger near the center of the cluster. The high concentration of GCs coinciding with ICL features at large cluster-centric distances in SMACS 0723 suggests that this stripping process need not occur only at the center of the potential. Relatedly, \citet{janssens17} find that the spatial distribution of ultra-compact dwarfs is strongly centrally concentrated in the Abell 2744 cluster, whereas ultra-diffuse galaxies have a flatter distribution. They hypothesize that diffuse galaxies may leave behind compact remnants as they dissolve in the cluster center. The central concentration we observe for dwarf galaxies seems to support this hypothesis, since we do not expect these galaxies to be able to retain their envelopes in the strong gravitational potential of the cluster center. 

Recent simulation work also implies that the spatial distribution of stellar light and GCs can be used to trace the dark matter halos of individual galaxies in addition to those of galaxy clusters \citep{reina-campos22,reina-campos23}. The present methodology produces individual cluster galaxy models that are conducive to such measurement, but we defer this investigation to future work. The CANUCS team is constructing detailed lens models for each cluster in the program in addition to the bCG and ICL models, which will facilitate the comparison of GC distributions with dark matter at the galaxy and cluster level.

\section{Summary and Conclusion}\label{sec:summary}

We have produced a model of the bright cluster galaxies and diffuse ICL in the \jwst ERO observations and archival \hst imaging of SMACS J0723.3-7327. The methodology presented here provides a significant increase in the available area in which we can perform photometry uncontaminated by the light of the lensing cluster, facilitating science goals both at the cluster redshift and beyond. We compare our models of the cluster galaxy and ICL to the spatial distribution of globular cluster (GC) and cluster dwarf galaxy candidates. Our main findings are summarized below.

\begin{enumerate}
    \item In general, the qualitative spatial distribution of compact sources traces galaxy light, but is more extended. Specifically, GC candidates are more closely associated with the diffuse ICL and cluster dwarf galaxy candidates are more centrally concentrated near the BCG. 

    \item We find a strong correlation between the surface density of each of our two samples with the surface brightness of total cluster light and quantify this relationship in the form of a linear sequence in log-log space. We perform a similar calculation for the relation between compact sources and the surface density of dark matter. 

    \item In agreement with previous work, we find that the cluster-centric density profile of GCs and ICL is steeper than that of dark matter. We obtain the same result for cluster dwarf galaxies. 
\end{enumerate}

Full cluster galaxy modeling of the five CANUCS lensing cluster fields will be completed in near future work. Applying the present analysis to this dataset will increase the number of intermediate redshift clusters with measured GC spatial distributions from two to seven and push the redshift boundary to $z \sim 0.5$. This will enable us to determine whether the correlations observed here hold in cluster environments of varying degrees of relaxation or in different mass regimes. In particular, the dramatic ICL features and their associated GCs separate from any member cluster galaxies in SMACS 0723 seem to potentially differ from other observed clusters. A larger sample of clusters will enable us to determine whether the trends we observe here hold generally, or if SMACS 0723 is somewhat anomalous in this regard.

\begin{acknowledgments}
NM acknowledges support from Canadian Space Agency grant 18JWST-GTO1. This research used the Canadian Advanced Network For Astronomy Research (CANFAR) operated in partnership by the Canadian Astronomy Data Centre and The Digital Research Alliance of Canada with support from the National Research Council of Canada the Canadian Space Agency, CANARIE and the Canadian Foundation for Innovation. NM, MB, AH and GR acknowledge support from the ERC Grant FIRSTLIGHT and from the Slovenian national research agency ARRS through grants N1-0238 and P1-0188. MB acknowledges support from the program HST-GO-16667, provided through a grant from the STScI under NASA contract NAS5-26555. 

This paper utilized data available on MAST under DOI \dataset[10.17909/XXXXX]{http://dx.doi.org/10.17909/XXXXXX}.
\end{acknowledgments}

\vspace{5mm}
\facilities{HST(ACS), \jwst(NIRISS,NIRCAM)}

\software{astropy \citep{price2018astropy}, photutils \citep{photutils20}}

\bibliography{refs}
\bibliographystyle{aasjournal}

\end{document}